\documentclass[
aps,prd,
12pt,
nopreprintnumbers,
showpacs,
eqsecnum,
nofootinbib
]{revtex4-1}

\usepackage{graphicx}
\usepackage{amssymb}
\usepackage{color}

\begin{document}

\title{DBI scalar f{}ield cosmology in $n\mathchar`-$DBI gravity}
\author{Nahomi Kan\footnote{Corresponding author}}\email[]{kan@gifu-nct.ac.jp}
\affiliation{National Institute of Technology, Gifu College,
Motosu-shi, Gifu 501-0495, Japan}
\author{Kiyoshi Shiraishi}\email[]{Deceased.}
\affiliation{
Graduate School of Sciences and Technology for Innovation, Yamaguchi
University, Yamaguchi-shi, Yamaguchi 753--8512, Japan}
\author{Maki Takeuchi}\email[]{maki\_t@yamaguchi-u.ac.jp}
\affiliation{
Graduate School of Sciences and Technology for Innovation, Yamaguchi
University, Yamaguchi-shi, Yamaguchi 753--8512, Japan}
\author{Mai Yashiki}\email[]{yashikimi@nbu.ac.jp}
\affiliation{Faculty of School of Engineering, Nippon Bunri University, 
Oita-shi, Oita 870-0397, Japan}

\date{\today}

\begin{abstract}
We study inflationary characterictics of the universe in $n$-DBI gravity, driven
by DBI-deformed scalar fields. In this paper, we consider the evolution of the
classical universe for a scalar potential whose equations of motion are expressed
by a BPS-like system of first-order differential equations. 
The advantage of a BPS-like system is that it allows the initial condition for the wave function of the Universe in minisuperspace quantum cosmology to be automatically determined by the Hamiltonian constraint.
By appropriately
choosing the prepotential underlying the potential, we can construct
single-scalar-field models that have the phases of exponential and power-law
expansions. We show that the slow-roll parameters for our models can be expressed
in terms of the prepotential with simple settings.
\end{abstract}

\maketitle

\section{Introduction}
\label{introduction}
The application of Dirac--Born--Infeld (DBI)-type scalar field theory to cosmology
has been considered for some time \cite{ST,AST}. The motivation was inspired by
$D$-brane physics, and early observations showed that there is an upper limit to
the speed of change of a uniform scalar field, and therefore slow-roll inflation
was expected. At the same time, we were also intrigued by the interesting property
that the equations of motion are reduced to a set of first-order differential
equations in the case of a special potential expressed by a prepotential, as in
the canonical scalar model \cite{SB,BGLM,BLRR,SA,Lidsey,Russo2}. Numerous studies
have shown that the simple DBI scalar model is likely to be ruled out by
future observational data \cite{MS}. However, it has been reported that
the mathematically-interesting special form of potentials emerges from $T\bar{T}$
deformations \cite{Taylor}, and that DBI-type kinetic terms can also be
obtained from canonical ones through $T\bar{T}$ deformations \cite{CINT}, so the
study of DBI-type scalar models and the associated set of special potentials is
quite valuable as equipment for theoretical arena.

On the other hand, DBI-type modified models of gravity theory have been
studied in a wide range of areas \cite{JHOR}. The $n$-DBI gravity model
\cite{HH,HHS,CHHS1,CHW,CHHS2}, which would fall into this category, is very
interesting, as it leads to a simple effective action for cosmological
models with flat space. 
The advantage of this model is that it leads to the equation of motion
that involves higer order terms in spatial derivatives, but in the cosmological
settings leads to a second order differential equation in time.
 Since this model is close to an extended
model of Ho\v{r}ava gravity \cite{Horava1,Horava2}, one should be careful to 
confirm the physical degrees of freedom of the graviton, but since it has a limit
that leads to Einstein gravity, it is thought that it can be adopted at least
for a model of the very early universe.

In this paper, we consider a flat isotropic universe accompanied by the evolution
of a DBI-type scalar field under extended $n$-DBI gravity. We can find a
special form of scalar potential and the evolution of the universe can be
described by a set of first-order differential equations such as the BPS
equations. The feature of this model is that, in the various limit of small
parameter functions, it can comprehensively describe systems with certain
combinations of Einstein gravity and canonical scalar fields, Einstein gravity
and DBI scalar fields, and $n$-DBI gravity and canonical scalar fields.

This paper is structured as follows. In the next section \ref{sec2}, we present
the Lagrangian of the model that we will consider. In section \ref{sec3}, we take
up a simple example for the prepotential of a single scalar field  and discuss
their cosmological development.  The slow-roll parameters associated with
inflation are denoted by the prepotential in each model with simple settings.
Slightly elaborated models of combinations of $n$-DBI gravity and the DBI scalar
theory are proposed and investigated in Sec.~\ref{sec4}. The last section is
devoted to a summary and future prospects.

We use metric signature $(-+\cdots+)$ and units $16\pi G=c=\hbar=1$. 
$\mu, \nu,\dots=0, 1,\cdots, D-1$ are coordinate indices of spacetime, while
$i, j=1, 2,\cdots, D-1$ are indices for space.

\section{Presentation of the model and effective action}
\label{sec2}

Our starting point is to give the following action for the DBI scalar model in
$n$-DBI gravity \cite{HH,HHS,CHHS1,CHW,CHHS2}:
\begin{eqnarray}
S&=&\int d^Dx \sqrt{-g}\left[\frac{1}{h(\phi)}
\sqrt{1+2
h(\phi)(R+\mathcal{K})}-\frac{1}{f(\phi)}\sqrt{1+f(\phi)g^{\mu\nu}G_{ab}(\phi)\partial_\mu\phi^a
\partial_\nu\phi^b}-V(\phi)
\right]\nonumber \\
&=&\int d^Dx \sqrt{-g}\left[\frac{1}{h(\phi)}
\left(\sqrt{1+2
h(\phi)(R+\mathcal{K})}-1\right)\right.\nonumber \\
&
&\qquad\qquad\qquad\left.-\frac{1}{f(\phi)}\left(\sqrt{1+f(\phi)g^{\mu\nu}G_{ab}(\phi)\partial_\mu\phi^a
\partial_\nu\phi^b}-1\right)-U(\phi)
\right]\,,
\end{eqnarray}
where $g$ denotes the determinant of the metric tensor $g_{\mu\nu}$,
$g^{\mu\nu}$ is the inverse of the metric tensor, 
$\phi^a$ $(a=1,\cdots, N)$ are the scalar fields, $f(\phi)$ and $h(\phi)$ are 
two functions of the scalar fields $\{\phi^a\}$, and
$U(\phi)=V(\phi)+f(\phi)^{-1}-h(\phi)^{-1}$ is the potential of scalar fields
$\{\phi^a\}$. The metric of the internal space
$G_{ab}(\phi)$ is given by a symmetric matrix,
$G_{ab}=G_{ba}$ and is generally dependent on scalar fields $\{\phi^a\}$.

Here, $R$ is the scalar curvature and $\mathcal{K}$ is defined as
\begin{equation}
\mathcal{K}\equiv -2\nabla_\mu(n^\mu\nabla_\nu n^\nu)\,,
\end{equation}
where $n^\mu$ is the unit timelike vector, which is firstly introduced in
the Arnowitt--Deser--Misner (ADM) formalism \cite{ADM1,ADM2}, and $\nabla_\mu$ is
the covariant derivative. It is known that the total derivative $\mathcal{K}$
is related to the Gibbons--Hawking--York term \cite{GH,York}.
Note that 
\begin{equation}
R+\mathcal{K}=G\equiv
g^{\mu\nu}(\Gamma^\rho_{\mu\sigma}\Gamma^\sigma_{\nu\rho}-
\Gamma^\rho_{\mu\nu}\Gamma^\sigma_{\sigma\rho})\,,
\end{equation}
which was described by Landau and Lifshitz \cite{LL}, where
$\Gamma^\lambda_{\mu\nu}$ is the Christoffel symbol. Notice also that the kinetic
term of scalar fields reduces to the one in the canonical scalar model in the
limit
$f(\phi)\rightarrow 0$, while the Einstein gravity is restored in the limit
$h(\phi)\rightarrow 0$.

Further, the metric of the $D$ dimensional spacetime is assumed as
\begin{equation}
ds^2=g_{\mu\nu}dx^\mu dx^\nu=-e^{2\gamma
\varphi(t)}dt^2+e^{2\beta\varphi(t)}\sum_{i=1}^{D-1} (dx^i)^2
=-e^{2\gamma
\varphi(t)}dt^2+a^2(t)\sum_{i=1}^{D-1} (dx^i)^2\,,
\end{equation}
where $\gamma$ is a constant, and
$\beta=\left(\sqrt{2(D-1)(D-2)}\right)^{-1}$,
and $a(t)$ is a scale factor of the flat, homogeneous and isotropic space.
Note that this metric becomes the standard
Friedmann--Lema\^{\i}tre--Robertson--Walker (FLRW) metric when $\gamma=0$, while
the metric takes the form of conformally flat when $\gamma=\beta$.

Assuming that all scalar fields are also spatially uniform, that is, a
function of time only, $\phi^a=\phi^a(t)$,
the effective Lagrangian on these variables is given by
\begin{equation}
L=e^{(2\gamma+\delta)\varphi}\left[\frac{1}{h(\phi)}\sqrt{1-
e^{-2\gamma\varphi}h(\phi)\dot{\varphi}^2}
-\frac{1}{f(\phi)}
\sqrt{1-{e^{-2\gamma\varphi}}f(\phi)G_{ab}(\phi)
\dot{\phi}^a\dot{\phi}^b}
-V(\phi)\right]\,,
\label{Lag1}
\end{equation}
where $\delta=(D-1)\beta-\gamma$ and 
the dot $(\dot{~})$ denotes the derivative with respect to time $t$.
Based on the Lagrangian (\ref{Lag1}), the conjugate momenta of $\varphi$ and
$\phi^a$ are expressed by
\begin{equation}
\Pi_\varphi=\frac{\partial
L}{\partial\dot{\varphi}}=\frac{-e^{\delta\varphi}\dot{\varphi}}{\sqrt{1-e^{-2\gamma\varphi}
h\dot{\varphi}^2}}\,,\quad\mbox{and}\quad
\Pi_a=\frac{\partial
L}{\partial\dot{\phi}^a}=\frac{e^{\delta\varphi}G_{ab}\dot{\phi}^b}{\sqrt{
1-{e^{-2\gamma\varphi}}fG_{cd}\dot{\phi}^c\dot{\phi}^d}}\,,
\label{mom3}
\end{equation}
respectively,
and the Hamiltonian $\mathcal{H}$ of the system can be found as
\begin{equation}
\mathcal{H}=e^{(2\gamma+\delta)\varphi}\left[-\frac{1}{h}
\sqrt{1+{e^{-2\alpha\varphi}}h\Pi_\varphi^2}
+\frac{1}{f}
\sqrt{1+{e^{-2\alpha\varphi}}fG^{ab}\Pi_a
\Pi_b}+V(\phi)\right]\,,
\end{equation}
where $G^{ab}$ is the inverse matrix of $G_{ab}$, and
\begin{equation}
\alpha=(D-1)\beta=\sqrt{\frac{D-1}{2(D-2)}}\,.
\end{equation}

Here, we assume the following two simultaneous equations:
\begin{equation}
\Pi_\varphi=-\epsilon\partial_\varphi\mathcal{W}(\varphi,\phi)\,,\quad
\Pi_a=-\epsilon \partial_a\mathcal{W}(\varphi,\phi)\,,
\label{bps3}
\end{equation}
where
\begin{equation}
\mathcal{W}(\varphi,\phi)=e^{\alpha\varphi}W(\phi)=a^{D-1}W(\phi)\,,
\end{equation}
and the constant $\epsilon$ satisfies $\epsilon^2=1$, 
$\partial_\varphi=\frac{\partial}{\partial\varphi}$, and
$\partial_a=\frac{\partial}{\partial\phi^a}$.

At the same time, if the potential takes the following form,
\begin{equation}
V(\phi)=
\frac{1}{h(\phi)}\sqrt{1+\alpha^2h(\phi)
W(\phi)^2}-\frac{1}{f(\phi)}\sqrt{1+f(\phi)G^{ab}(\phi)\partial_a
{W(\phi)}\partial_b{W(\phi)}}\,,
\end{equation}
the Hamiltonian constraint%
\footnote{As is well known, the Hamiltonian constraint is derived by,
replacing $t\rightarrow Nt$ in the action $S=\int L dt$ and regarding that the
variation
$\frac{\delta S}{\delta N}$ vanishes \cite{ADM1,ADM2}.}
 $\mathcal{H}=0$ is classically satisfied by taking
(\ref{bps3}) with (\ref{mom3}).
Since the scalar potential $V(\phi)$ or $U(\phi)$ is controlled by $W(\phi)$, 
we call
$W(\phi)$ prepotential.

One can find the equations of motion for the system are
\begin{eqnarray}
\dot{\Pi}_\varphi
&=&e^{(2\gamma+\delta)\varphi}
\left[\frac{\gamma}{h}
\sqrt{1+{e^{-2\alpha\varphi}}h\Pi_\varphi^2}
-\frac{\gamma}{f}
\sqrt{1+{e^{-2\alpha\varphi}}fG^{ab}
\Pi_a\Pi_b}\right.\nonumber \\
& &\qquad\qquad\left.+\frac{\alpha}{h}
\frac{1}{\sqrt{1+{e^{-2\alpha\varphi}}h\Pi_\varphi^2}}
-\frac{\alpha}{f}
\frac{1}{\sqrt{1+{e^{-2\alpha\varphi}}fG^{ab}
\Pi_a\Pi_b}}-(2\gamma+\delta)
V\right]\,,
\label{eomv}
\end{eqnarray}
\begin{eqnarray}
\dot{\Pi}_a
&=&-e^{(2\gamma+\delta)\varphi}\partial_aV
+\frac{e^{-\delta\varphi}}{2}\frac{(\partial_aG^{bc})
\Pi_b\Pi_c}{\sqrt{1+{e^{-2\alpha\varphi}}fG^{de}
\Pi_d\Pi_e}}\nonumber \\
& &
-e^{(2\gamma+\delta)\varphi}\frac{\partial_ah}{2h^2}\left[
\sqrt{1+{e^{-2\alpha\varphi}}h\Pi_\varphi^2}+
\frac{1}{\sqrt{1+{e^{-2\alpha\varphi}}h\Pi_\varphi^2}}\right]\nonumber
\\ & &
+e^{(2\gamma+\delta)\varphi}\frac{\partial_af}{2f^2}\left[
\sqrt{1+{e^{-2\alpha\varphi}}fG^{ab}\Pi_a\Pi_b}+
\frac{1}{\sqrt{1+{e^{-2\alpha\varphi}}fG^{ab}\Pi_a\Pi_b}}\right]\,.
\label{eomp}
\end{eqnarray}
One can confirm that these equations hold if the BPS-like equations (\ref{bps3})
with (\ref{mom3}) are substituted, noting that
\begin{eqnarray}
& &\sqrt{1-{e^{-2\gamma\varphi}}h\dot{\varphi}^2}\sqrt{1+{e^{-2\alpha\varphi}}
h(\partial_\varphi\mathcal{W})^2}\nonumber \\
&=&
\sqrt{1-{e^{-2\gamma\varphi}}fG_{ab}
\dot{\phi}^a\dot{\phi}^b}\sqrt{1+{e^{-2\alpha\varphi}}fG^{ab}\partial_a
\mathcal{W}\partial_b\mathcal{W}}=1\,,
\label{id}
\end{eqnarray}
and notice that $\partial_a G^{bc}=-G^{bd}G^{ce}\partial_aG_{de}$.

Notice that the Lagrangian (\ref{Lag1}) can be rewritten in the form
\begin{eqnarray}
L&=&-\frac{1}{2}e^{-\delta\varphi}\sqrt{1-{e^{-2\gamma\varphi}}h\dot{\varphi}^2}
\left[\left(\sqrt{1-{e^{-2\gamma\varphi}}h\dot{\varphi}^2}\right)^{-1}
e^{\delta\varphi}\dot{\varphi}
-\epsilon\partial_\varphi\mathcal{W}\right]^2\nonumber \\
& &+\frac{1}{2}e^{-\delta\varphi}\sqrt{1-{e^{-2\gamma\varphi}}fG_{ab}
\dot{\phi}^a\dot{\phi}^b}\,G_{cd}\left[\left(\sqrt{1-{e^{-2\gamma\varphi}}fG_{ab}
\dot{\phi}^a\dot{\phi}^b}\right)^{-1}e^{\delta\varphi}\dot{\phi}^c
+\epsilon G^{ce}\partial_e\mathcal{W}\right]\nonumber \\
& &\qquad\qquad\qquad\times
\left[\left(\sqrt{1-{e^{-2\gamma\varphi}}fG_{ab}
\dot{\phi}^a\dot{\phi}^b}\right)^{-1}e^{\delta\varphi}\dot{\phi}^d
+\epsilon G^{de}\partial_e\mathcal{W}\right]\nonumber \\
& &+\frac{1}{2}\frac{e^{(2\gamma+\delta)\varphi}}{h}
\left(\sqrt{1-{e^{-2\gamma\varphi}}h\dot{\varphi}^2}\right)^{-1}\left[
\sqrt{1-{e^{-2\gamma\varphi}}h\dot{\varphi}^2}\sqrt{1+{e^{-2\alpha\varphi}}h
(\partial_\varphi\mathcal{W})^2}-1\right]^2\nonumber \\
& &-\frac{1}{2}\frac{e^{(2\gamma+\delta)\varphi}}{f}
\left(\sqrt{1-{e^{-2\gamma\varphi}}fG_{ab}
\dot{\phi}^a\dot{\phi}^b}\right)^{-1}\nonumber \\
& &\qquad\qquad\qquad\times\left[
\sqrt{1-{e^{-2\gamma\varphi}}fG_{ab}
\dot{\phi}^a\dot{\phi}^b}\sqrt{1+{e^{-2\alpha\varphi}}fG^{ab}\partial_a
\mathcal{W}\partial_b\mathcal{W}}-1\right]^2\nonumber \\
& &-\epsilon(\dot{\varphi}\partial_\varphi\mathcal{W}+
\dot{\phi}^a\partial_a\mathcal{W})\,,
\end{eqnarray}
so, it is apparent that the equations (\ref{bps3}) represent a stationary point of
the action.

\section{simple examples and slow-roll parameters}
\label{sec3}

Here, we consider the single scalar case (i.e., $G_{ab}\rightarrow 1$), for
simplicity. Moreover, we here set $\gamma=0$ to obtain the standard FLRW metric.
The BPS-like equations then becomes
\begin{equation}
\dot{\varphi}=\frac{\alpha W(\phi)}{\sqrt{1+\alpha^2h(\phi)W(\phi)^2}}\,,\quad
\dot{\phi}=-\frac{W'(\phi)}{\sqrt{1+f(\phi)W'(\phi)^2}}\,,
\end{equation}
where we have chosen $\epsilon=1$, for it leads to an expanding universe for
$W>0$.

Note that the e-fold number $N$ between $t=t_i$ and $t=t_f$ is expressed in
terms of
$\varphi$, as
\begin{equation}
N={\beta}(\varphi(t_f)-\varphi(t_i))\,.
\end{equation}

Before we get into the individual examples, we consider {the special
case with constant
$f(\phi)=f_0$ and
$h(\phi)=h_0$}. In this case, the slow-roll parameters are expressed by the
single field prepotential
$W(\phi)$, after some calculations, as
\begin{eqnarray}
\varepsilon_H&=&-\frac{\dot{H}}{H^2}=\frac{2(D-2)}{\sqrt{1+f_0W'(\phi)^2}
\sqrt{1+\alpha^2h_0W(\phi)^2}}\frac{W'(\phi)^2}{W(\phi)^2}\,,\\
\eta_H&=&-\frac{\dot{u}}{Hu}=\frac{2(D-2)\sqrt{1+\alpha^2h_0W(\phi)^2}}{\sqrt{1+f_0W'(\phi)^2}
}\frac{W''(\phi)}{W(\phi)}\,.
\end{eqnarray}
These are valid for a general function for $W(\phi)$.
Here, $u$ is defined by $\frac{\dot{\phi}}{\sqrt{1-f_0\dot{\phi}^2}}$,
which satisfies $\dot{u}+(D-1)Hu+V'(\phi)=0$.
We should study the behavior related to the inflationary scenario more deeply
in the future, but we should remark that the correction factor
$\sqrt{1+\alpha^2h_0W(\phi)^2}$ from $n$-DBI gravity acts in opposite direction
in correcting the two slow-roll parameters.
Therefore in this case, if the observations constrain both slow-roll parameters to
be very small, the absolute value of $h_0$ may be limited to a very small value.

In the following discussion of
{slow-roll parameters}, we will always assume {constants $f_0$
and
$h_0$}.

\subsection{the exponential prepotential}

We assume exponential forms for $W(\phi)$, $f(\phi)$, and $h(\phi)$, i.e.,
\begin{equation}
W(\phi)=W_\lambda e^{-\lambda\phi}\,,\quad
f(\phi)=f_\mu e^{-\mu\phi}\,,\quad
h(\phi)=h_\kappa e^{-\kappa\phi}\,,
\end{equation}
where $W_\lambda$, $f_\mu$, $h_\kappa$, $\lambda$, $\mu$, and $\kappa$ are
constant. Note that the exponential functions naturally arise from
compactifications of field theories and string theories.
In the canonical Einstein-scalar theory (as obtained in the limit
$f(\phi)=h(\phi)=0$), we expect that both
$\varphi(t)$ and
$\phi(t)$ are monotonically increasing functions. The BPS-like equations in the
present case are further simplified as
\begin{equation}
\dot{\varphi}=\frac{\alpha
W_\lambda e^{-\lambda\phi}}{\sqrt{1+\alpha^2h_\kappa
W_\lambda^2e^{-(\kappa+2\lambda)\phi}}}\,,\quad
\dot{\phi}=\frac{\lambda W_\lambda
e^{-\lambda\phi}}{\sqrt{1+\lambda^2f_\mu W_\lambda^2 e^{-(\mu+2\lambda)\phi}}}\,.
\end{equation}
Accordingly, we also find
\begin{equation}
\frac{d\varphi}{d\phi}=\frac{\alpha\sqrt{1+\lambda^2f_\mu W_\lambda^2
e^{-(\mu+2\lambda)\phi}}}{\lambda\sqrt{1+\alpha^2h_\kappa
W_\lambda^2e^{-(\kappa+2\lambda)\phi}}}\,.
\label{pp}
\end{equation}
In this case, the potential becomes
\begin{equation}
V(\phi)=\frac{e^{\kappa\phi}}{h_\kappa}\sqrt{1+\alpha^2h_\kappa
W_\lambda^2e^{-(\kappa+2\lambda)\phi}}-\frac{e^{\mu\phi}}{f_\mu}\sqrt{1+\lambda^2
f_\mu W_\lambda^2e^{-(\mu+2\lambda)\phi}}\,.
\end{equation}

Below, we consider parameter choices that produce some characteristic cases.

\subsubsection{$\lambda=0$}
In this case, the prepotential $W(\phi)$ is a constant, $W_\lambda$.
We find the solutions
\begin{equation}
H=\frac{\dot{a}}{a}=\beta\dot{\varphi}=\frac{\alpha\beta
W_\lambda}{\sqrt{1+\alpha^2h_\kappa W_\lambda^2e^{-\kappa\phi_0}}}\,,\quad
{\phi}=\phi_0=\mbox{constant}\,.
\end{equation}
Then, the universe undergoes exact de Sitter expansion. 

\subsubsection{In the region of $\lambda\phi\gg 1$, for $\mu>0$ and $\kappa>0$}
In the asymptotic region of $\lambda\phi\gg 1$, which may correspond to the
late-time expansion, we obtained the approximate solution
\begin{equation}
e^{\alpha\varphi}=a^{D-1}\propto(t-t_1)^{\frac{\alpha^2}{\lambda^2}}\,,\quad
\phi\approx \frac{1}{\lambda}\ln[\lambda^2W_\lambda(t-t_1)]\,,
\label{as}
\end{equation}
where $t_1$ is a constant.
The cosmological model of a scalar field with an exponential
potential has been studied by many authors until \cite{Russo0,Neupane,ENO}. 
The asymptotic solution (\ref{as}) coincides with their solution. 
Note that, from (\ref{pp}),
$\varphi(t_f)-\varphi(t_i)-\frac{\alpha}{\lambda}(\phi(t_f)-\phi(t_i))\approx0$
for large $\phi$.

\subsubsection{$\mu=-2\lambda$ and $\kappa=-2\lambda$}
In this case,
\begin{equation}
e^{\alpha\varphi}=a^{D-1}\propto(t-t_1)^{\frac{\alpha^2\sqrt{1+\lambda^2f_\mu
W_\lambda^2 }}{\lambda^2\sqrt{1+\alpha^2h_\kappa W_\lambda^2}}}\,,\quad
\phi\approx
\frac{1}{\lambda}\ln\left[\frac{\lambda^2W_\lambda}{\sqrt{1+\lambda^2f_\mu
W_\lambda^2}}(t-t_1)
\right]\,,
\end{equation}
where $t_1$ is a constant.
It is interesting that the index of the power-law expansion is modified
from the previous case. Notice that since $h$ and $f$ increase as $\phi$
increases if $\lambda>0$, the model moves away from the canonical Einstein-scalar
theory asymptotically with increasing $\phi$.%
\footnote{Therefore, one of interesting subjects to study is the possibility of
the time-varying parameters (especially for $\kappa$ and $\mu$).} Incidentally, we
find that the scalar potential takes a simple exponential form in this case, i.e.,
\begin{equation}
V(\phi)=\left[\frac{1}{h_\kappa}\sqrt{1+\alpha^2h_\kappa
W_\lambda^2}-\frac{1}{f_\mu}\sqrt{1+\lambda^2
f_\mu W_\lambda^2}\right]e^{-2\lambda\phi}\,,
\end{equation}
and $U(\phi)=V(\phi)+(f_\mu^{-1}-h_\kappa^{-1})e^{-2\lambda\phi}$.

\subsubsection{$h_\kappa=0$}

In this case, the relation of $\varphi$ and $\phi$ can be solved analytically.
That is,
\begin{eqnarray}
& &\varphi(t_f)-\varphi(t_i)-\frac{\alpha}{\lambda}(\phi(t_f)-\phi(t_i))\nonumber
\\ &=&-\frac{2\alpha}{\lambda(\mu+2\lambda)}
\left[\sqrt{1+\lambda^2f_\mu W_\lambda^2
e^{-(\mu+2\lambda)\phi}}-1-\ln\left[\frac{1+\sqrt{1+\lambda^2f_\mu W_\lambda^2
e^{-(\mu+2\lambda)\phi}}}{2}\right]\right]^{\phi=\phi(t_f)}_{\phi=\phi(t_i)}\,.
\end{eqnarray}
This value becomes negative for $\lambda>0$, $\mu+2\lambda>0$, and $f_\mu>0$.
Thus, the power-law expansion becomes supressed for positive $f_\mu$ and enhanced
for negative $f_\mu$ under the same change of the value of $\phi$.

\subsubsection{$f_\mu=0$}
Also in this case, the relation of $\varphi$ and $\phi$ can be solved
analytically:
\begin{equation}
\varphi(t_f)-\varphi(t_i)-\frac{\alpha}{\lambda}(\phi(t_f)-\phi(t_i))=
\frac{2\alpha}{\lambda(\kappa+2\lambda)}
\left[\ln\left[\frac{1+\sqrt{1+\alpha^2h_\kappa W_\lambda^2
e^{-(\kappa+2\lambda)\phi}}}{2}\right]\right]^{\phi=\phi(t_f)}_{\phi=\phi(t_i)}\,.
\end{equation}
This value becomes positive for $\lambda>0$, $\kappa+2\lambda>0$, and
$h_\kappa>0$. Thus, the power-law expansion becomes enhanced for positive
$h_\kappa$ and supressed for negative $h_\kappa$ under the same change of the
value of
$\phi$.

\subsubsection{$\lambda^2f_\mu=\alpha^2h_\kappa$ and $\mu=\kappa$}
In this case,
$\varphi(t_f)-\varphi(t_i)=\frac{\alpha}{\lambda}(\phi(t_f)-\phi(t_i))$ exactly,
which is recognized from (\ref{pp}). Incidentally, $V(\phi)$ takes the form
\begin{equation}
V(\phi)=\frac{1}{h_\kappa}\left(1-\frac{\lambda^2}{\alpha^2}\right)e^{\kappa\phi}
\sqrt{1+\alpha^2h_\kappa
W_\lambda^2e^{-(\kappa+2\lambda)\phi}}\,.
\end{equation}

\subsubsection{slow-roll parameters for $f=f_0$ and $h=h_0$ (or, $\mu=\kappa=0$
and $f_\mu=f_0$ and $h_\kappa=h_0$)}
If $W(\phi)=W_\lambda e^{-\lambda\phi}$, we find
\begin{eqnarray}
\varepsilon_H&=&\frac{2(D-2)\lambda^2}
{\sqrt{1+\lambda^2f_0W_\lambda^2e^{-2\lambda\phi}}
\sqrt{1+\alpha^2h_0W_\lambda^2e^{-2\lambda\phi}}}\,,\\
\eta_H&=&\frac{2(D-2)\lambda^2
\sqrt{1+\alpha^2h_0W_\lambda^2e^{-2\lambda\phi}}}
{\sqrt{1+\lambda^2f_0W_\lambda^2e^{-2\lambda\phi}}}\,.
\label{srp1}
\end{eqnarray}
Therefore, to obtain small values for the parameters ($\varepsilon_H, \eta_H\ll
1$ for an early time ($\lambda\phi\sim 0$)), a reasonably small $\lambda$ is
required, otherwise $\sqrt{1+\alpha^2h_0W_\lambda^2}\gg 1$ and
$\sqrt{1+\alpha^2h_0W_\lambda^2}\ll
\sqrt{1+\lambda^2f_0W_\lambda^2}$.

\subsection{the quadratic prepotential}
Here, we assume
\begin{equation}
W(\phi)=g_0+g\phi^2\,,
\end{equation}
where $g_0$ and $g$ are constant. For simplicity, we also assume that
$f(\phi)=f_0$ and $h(\phi)=h_0$ are constant.
Then, the expansion rate becomes
\begin{equation}
H=\frac{\dot{a}}{a}=\beta\dot{\varphi}=\frac{\alpha\beta
(g_0+g\phi^2)}{\sqrt{1+\alpha^2h_0(g_0+g\phi^2)^2}}\,.
\end{equation}
Thus, the time variation of $\phi$ is small, $H$ becomes nearly constant.
The slow-roll parameter $\varepsilon_H$ reads
\begin{equation}
\varepsilon_H=\frac{2(D-2)}{\sqrt{1+4f_0g^2\phi^2}
\sqrt{1+\alpha^2h_0(g_0+g\phi^2)^2}}\frac{4g^2\phi^2}{(g_0+g\phi^2)^2}\,,\\
\end{equation}
and thus for large $\phi$, $\varepsilon_H\ll 1$. Especially, if $\phi$ is larger
than
$\sqrt{g_0/g}$, $(2\sqrt{f_0}g)^{-1}$, and $(\alpha\sqrt{h_0}g)^{-1/2}$,
$\varepsilon_H\sim 4(D-2)/(\sqrt{f_0h_0}\alpha g^2\phi^5)\ll 1$.
On the other hand, the slow-roll parameter $\eta_H$ becomes
\begin{equation}
\eta_H=\frac{2(D-2)\sqrt{1+\alpha^2h_0(g_0+g\phi^2)^2}}{\sqrt{1+4f_0g^2\phi^2}
}\frac{2g}{g_0+g\phi^2}\,.
\end{equation}
If the value of $\phi$ is very large as previously assumed, $\eta_H\sim 
2(D-2)\alpha\sqrt{h_0}/(\sqrt{f_0}\phi)$, which is independent of the coefficient
$g$.

\subsection{the reciprocal cosh prepotential}
Here, we assume 
\begin{equation}
W(\phi)=w^2 [\cosh(m\phi)]^{-1}\,,
\end{equation}
where $w$ and $m$ are constant.
This prepotential was also tested in our previous paper \cite{KSTY}.
Because this prepotential in the region $m\phi\gg 1$ can be approximated by the
exponential function, we consider the region $m\phi\ll 1$ here.
The parameter $\varepsilon_H$ can be arbitrarily small, since $W'(\phi)\ll 1$
when $m\phi\ll 1$. On the other hand, the slow-roll parameter $\eta_H$
approaches, in the limit $m\phi\rightarrow 0$,
\begin{equation}
\eta_H=-2(D-2)m^2\sqrt{1+\alpha^2h_0w^4}\,,
\end{equation}
for $f(\phi)=f_0$ and $h(\phi)=h_0$ are constant.
To obtain small $\eta_H$, we consider small $m$ as well as negative values for
$h_0$, which is the characteristic parameter of the $n$-DBI gravity.

\section{the other models}
\label{sec4}

\subsection{the other model (A)}

One of the other models is  the following combined model of the DBI scalar theory
and
$n$-DBI gravity:
\begin{eqnarray}
S&=&\int d^Dx \sqrt{-g}\left[\frac{1}{h(\phi)}
\sqrt{1+
h(\phi)(2[R+\mathcal{K}]-g^{\mu\nu}G_{ab}(\phi)\partial_\mu\phi^a
\partial_\nu\phi^b)}-V(\phi)
\right]\nonumber \\
&=&\int d^Dx \sqrt{-g}\left[\frac{1}{h(\phi)}
\left(\sqrt{1+
h(\phi)(2[R+\mathcal{K}]-g^{\mu\nu}G_{ab}(\phi)\partial_\mu\phi^a
\partial_\nu\phi^b)}-1\right)-U(\phi)
\right]\,,
\end{eqnarray}
where 
$U(\phi)=V(\phi)-h(\phi)^{-1}$. Obviously, the limit $h(\phi)\rightarrow 0$ again
yields the canonical scalar theory in the Einstein gravity.
The effective action is found to be
\begin{equation}
L=e^{(2\gamma+\delta)\varphi}\left[
\frac{1}{h(\phi)}
\sqrt{1-{e^{-2\gamma\varphi}}h(\phi)(\dot{\varphi}^2-G_{ab}(\phi)
\dot{\phi}^a\dot{\phi}^b)}
-V(\phi)\right]\,,
\end{equation}
and thus the conjugate momenta are
\begin{equation}
\Pi_\varphi=\frac{\partial
L}{\partial\dot{\varphi}}=\frac{-e^{\delta\varphi}\dot{\varphi}}
{\sqrt{1-{e^{-2\gamma\varphi}}h(\dot{\varphi}^2-G_{ab}
\dot{\phi}^a\dot{\phi}^b)}}\,,\quad
\Pi_a=\frac{\partial
L}{\partial\dot{\phi}^a}=\frac{e^{\delta\varphi}G_{ab}\dot{\phi}^b}
{\sqrt{1-{e^{-2\gamma\varphi}}h(\dot{\varphi}^2-G_{ab}
\dot{\phi}^a\dot{\phi}^b)}}\,,
\label{mom2}
\end{equation}
respectively,
and the Hamiltonian can be found as
\begin{equation}
\mathcal{H}=e^{(2\gamma+\delta)\varphi}\left[-\frac{1}{h(\phi)}
\sqrt{1+{e^{-2\alpha\varphi}}h(\phi)(\Pi_\varphi^2-G^{ab}(\phi)\Pi_a
\Pi_b)}+V(\phi)\right]\,.
\end{equation}

Here, we assume the following two simultaneous equations:
\begin{equation}
\Pi_\varphi=-\epsilon\partial_\varphi\mathcal{W}(\varphi,\phi)\,,\quad
\Pi_a=-\epsilon \partial_a\mathcal{W}(\varphi,\phi)\,,
\label{bps2}
\end{equation}
where
$\mathcal{W}(\varphi,\phi)=e^{\alpha\varphi}W(\phi)=a^{D-1}W(\phi)$.

At the same time, if the potential takes the following form,
\begin{equation}
V(\phi)=
\frac{1}{h(\phi)}\sqrt{1+h(\phi)[\alpha^2W(\phi)^2-G^{ab}(\phi)\partial_a
{W(\phi)}\partial_b{W(\phi)}]}\,,
\end{equation}
the classical Hamiltonian constraint $\mathcal{H}=0$ is satisfied by taking
(\ref{bps2}) with (\ref{mom2}).

The equations of motion for the system are
\begin{eqnarray}
\dot{\Pi}_\varphi
&=&e^{(2\gamma+\delta)\varphi}
\left[
\frac{\gamma}{h}
\sqrt{1+{e^{-2\alpha\varphi}}h[\Pi_\varphi^2-G^{ab}
\Pi_a\Pi_b]}\right.\nonumber \\
& &\qquad\qquad\left.
+\frac{\alpha}{h}
\frac{1}{\sqrt{1+{e^{-2\alpha\varphi}}h[\Pi_\varphi^2-G^{ab}
\Pi_a\Pi_b]}}-(2\gamma+\delta)
V\right]\,,
\label{eomv}
\end{eqnarray}
\begin{eqnarray}
\dot{\Pi}_a
&=&-e^{(2\gamma+\delta)\varphi}\partial_aV
+\frac{e^{-\delta\varphi}}{2}\frac{(\partial_aG^{bc})
\Pi_b\Pi_c}{\sqrt{1+{e^{-2\alpha\varphi}}h[\Pi_\varphi^2-G^{ab}
\Pi_a\Pi_b]}}\nonumber \\
& &
-e^{(2\gamma+\delta)\varphi}\frac{\partial_ah}{2h^2}\left[
\sqrt{1+{e^{-2\alpha\varphi}}h[\Pi_\varphi^2-G^{ab}
\Pi_a\Pi_b]}+
\frac{1}{\sqrt{1+{e^{-2\alpha\varphi}}h[\Pi_\varphi^2-G^{ab}
\Pi_a\Pi_b]}}\right]\,.
\label{eomp}
\end{eqnarray}
One can confirm that these equations hold if the BPS-like equations (\ref{bps2})
with (\ref{mom2}) are substituted, noting that
\begin{equation}
\sqrt{1-{e^{-2\gamma\varphi}}h[\dot{\varphi}^2-G_{ab}
\dot{\phi}^a\dot{\phi}^b]}\sqrt{1+{e^{-2\alpha\varphi}}h[
(\partial_\varphi\mathcal{W})^2-G^{ab}\partial_a
\mathcal{W}\partial_b\mathcal{W}]}=1\,.
\end{equation}
The BPS-like equations is a stationary point of the action, since
the effective Lagrangian can be rewritten as
\begin{eqnarray}
L&=&\frac{1}{2}e^{-\delta\varphi}\sqrt{1-{e^{-2\gamma\varphi}}h(\dot{\varphi}^2-G_{ab}
\dot{\phi}^a\dot{\phi}^b)}\nonumber
\\
&
&\quad\times\left\{
G_{cd}\left[\left(\sqrt{1-{e^{-2\gamma\varphi}}h[\dot{\varphi}^2-G_{ab}
\dot{\phi}^a\dot{\phi}^b]}\right)^{-1}e^{\delta\varphi}\dot{\phi}^c
+\epsilon G^{ce}\partial_e\mathcal{W}\right]\right.\nonumber \\
& &\qquad\qquad\qquad\times
\left[\left(\sqrt{1-{e^{-2\gamma\varphi}}h[\dot{\varphi}^2-G_{ab}
\dot{\phi}^a\dot{\phi}^b]}\right)^{-1}e^{\delta\varphi}\dot{\phi}^d
+\epsilon G^{de}\partial_e\mathcal{W}\right]\nonumber \\
& &\left.\qquad-\left[\left(\sqrt{1-{e^{-2\gamma\varphi}}h[\dot{\varphi}^2-G_{ab}
\dot{\phi}^a\dot{\phi}^b]}\right)^{-1}e^{\delta\varphi}\dot{\varphi}-\epsilon\partial_\varphi
\mathcal{W}\right]^2\right\}
\nonumber
\\& &+\frac{1}{2}\frac{e^{(2\gamma+\delta)\varphi}}{h}
\left(\sqrt{1-{e^{-2\gamma\varphi}}h[\dot{\varphi}^2-G_{ab}
\dot{\phi}^a\dot{\phi}^b]}\right)^{-1}\nonumber \\
& &\times\left[
\sqrt{1-{e^{-2\gamma\varphi}}h[\dot{\varphi}^2-G_{ab}
\dot{\phi}^a\dot{\phi}^b]}\sqrt{1+{e^{-2\alpha\varphi}}h[
(\partial_\varphi\mathcal{W})^2-G^{ab}\partial_a
\mathcal{W}\partial_b\mathcal{W}]}-1\right]^2\nonumber \\
& &-\epsilon(\dot{\varphi}\partial_\varphi\mathcal{W}+
\dot{\phi}^a\partial_a\mathcal{W})\,.
\end{eqnarray}

In this model, two slow-roll parameters are expressed, when we consider the
single-field case and $h=h_0$ is constant, as
\begin{eqnarray}
\varepsilon_H&=&-\frac{\dot{H}}{H^2}=2(D-2)\frac{1+
h_0(W(\phi)W''(\phi)-W'(\phi)^2)}{1+
h_0(\alpha^2W(\phi)^2-W'(\phi)^2)}\frac{W'(\phi)^2}{W(\phi)^2}\,,\\
\eta_H&=&-\frac{\dot{u}}{Hu}=2(D-2)
\frac{W''(\phi)}{W(\phi)}\,.
\label{srp2}
\end{eqnarray}
We should note that $\eta_H$ is not corrected by $h_0$.
Incidentally, for the case with $W(\phi)=W_\lambda e^{-\lambda\phi}$ as previously
treated, we find
\begin{equation}
\varepsilon_H=\frac{2(D-2)\lambda^2}{1+
h_0(\alpha^2-\lambda^2)W_\lambda^2e^{-2\lambda\phi}}\,,\quad
\eta_H=2(D-2)\lambda^2\,.
\end{equation}
In this case, the constant $\lambda$ must be small in order for $\varepsilon_H$
and $\eta_H$ to be small, and unfortunately the characteristics of the DBI-type
model do not appear. 

For quadratic prepotential $W(\phi)=g_0+g\phi^2$, we find small $\varepsilon_H$
and $\eta_H$ for large $\phi$. On the other hand, for the reciprocal cosh
prepotential
$W(\phi)=w^2[\cosh(m\phi)]^{-1}$ with small $\phi$, small $\varepsilon_H$
and $\eta_H$ are obtained if $m^2\ll 1$.

\subsection{the other model (B)}

Another (and the last) model is described by the following action:
\begin{eqnarray}
& &S=\int d^Dx \sqrt{-g}\left[\frac{1}{h(\phi)}
\left(\sqrt{1-\alpha^2h(\phi)W(\phi)^2}\sqrt{1+
2h(\phi)[R+\mathcal{K}]}-1\right)
\right.\nonumber \\
& &-\frac{1}{f(\phi)}\left.
\left(\sqrt{1-f(\phi)G^{ab}(\phi)\partial_aW(\phi)
\partial_bW(\phi)}\sqrt{1+f(\phi)g^{\mu\nu}G_{ab}(\phi)\partial_\mu\phi^a
\partial_\nu\phi^b)}-1\right)
\right]\,.
\end{eqnarray}
The effective Lagrangian derived from the action is
\begin{eqnarray}
& &L=e^{(2\gamma+\delta)\varphi}\left[
\frac{1}{h(\phi)}\left(\sqrt{1-\alpha^2h(\phi)W(\phi)^2}
\sqrt{1-{e^{-2\gamma\varphi}}h(\phi)\dot{\varphi}^2}
-1\right)\right.\nonumber \\
& &\left.
-\frac{1}{f(\phi)}\left(\sqrt{1-f(\phi)G^{ab}\partial_aW(\phi)
\partial_bW(\phi)}
\sqrt{1-{e^{-2\gamma\varphi}}f(\phi)G_{ab}(\phi)
\dot{\phi}^a\dot{\phi}^b}
-1\right)\right]\,.
\end{eqnarray}
Then the conjugate momenta are
\begin{equation}
\Pi_\varphi=\frac{\partial
L}{\partial\dot{\varphi}}=\frac{-e^{\delta\varphi}\dot{\varphi}\sqrt{1-\alpha^2h
W^2}}
{\sqrt{1-{e^{-2\gamma\varphi}}h\dot{\varphi}^2}}\,,\quad
\Pi_a=\frac{\partial
L}{\partial\dot{\phi}^a}=\frac{e^{\delta\varphi}G_{ab}\dot{\phi}^b\sqrt{1-f
G^{ab}\partial_aW
\partial_bW}}
{\sqrt{1-{e^{-2\gamma\varphi}}fG_{ab}
\dot{\phi}^a\dot{\phi}^b}}\,,
\label{lastP}
\end{equation}
and the Hamiltonian is found to be
\begin{eqnarray}
\mathcal{H}&=&e^{(2\gamma+\delta)\varphi}\left[-\frac{1}{h(\phi)}
\left(\sqrt{1+h(\phi)(e^{-2\alpha\varphi}\Pi_\varphi^2-\alpha^2W(\phi)^2)}-1\right)\right.\nonumber
\\ & &\left.+\frac{1}{f(\phi)}
\left(\sqrt{1+f(\phi)G^{ab}(\phi)(e^{-2\alpha\varphi}\Pi_a\Pi_b-
\partial_a W(\phi)\partial_b W(\phi))}-1\right)\right]\,.
\end{eqnarray}

If the following equations hold, $\mathcal{H}=0$, classically:
\begin{equation}
\Pi_\varphi=-\epsilon\partial_\varphi\mathcal{W}(\varphi,\phi)\,,\quad
\Pi_a=-\epsilon \partial_a\mathcal{W}(\varphi,\phi)\,,
\label{BPSlast}
\end{equation}
where
$\mathcal{W}(\varphi,\phi)=e^{\alpha\varphi}W(\phi)=a^{D-1}W(\phi)$.
Although we no longer write out the equations of motion here, the eqautions of
motion are satisfied if these BPS-type equations are satisfied. 
The equations (\ref{BPSlast}) with (\ref{lastP}) lead to
\begin{eqnarray}
\sqrt{1-{e^{-2\gamma\varphi}}h(\phi)\dot{\varphi}^2}
&=&\sqrt{1-\alpha^2h(\phi)W(\phi)^2}\,,\label{eq1}
 \\
\sqrt{1-{e^{-2\gamma\varphi}}f(\phi)G_{ab}(\phi)
\dot{\phi}^a\dot{\phi}^b}&=&\sqrt{1-f(\phi)G^{ab}\partial_aW(\phi)\partial_bW(\phi)}\,.
\label{eq2}
\end{eqnarray}
Also, the effective Lagrangian can be rewritten as
\begin{eqnarray}
L&=&-\frac{1}{2}e^{-\delta\varphi}
\sqrt{1-\alpha^2hW^2}
\sqrt{1-{e^{-2\gamma\varphi}}h\dot{\varphi}^2}\nonumber \\
& &\quad\times
\left[\left(\sqrt{1-{e^{-2\gamma\varphi}}h\dot{\varphi}^2}\right)^{-1}
e^{\delta\varphi}\dot{\varphi}-\left(\sqrt{1-\alpha^2hW^2}\right)^{-1}
\epsilon\partial_\varphi
\mathcal{W}\right]^2\nonumber\\
& &+\frac{1}{2h}e^{(2\gamma+\delta)\varphi}
\sqrt{1-\alpha^2hW^2}
\sqrt{1-{e^{-2\gamma\varphi}}h\dot{\varphi}^2}\left[
\left(\sqrt{1-{e^{-2\gamma\varphi}}
h\dot{\varphi}^2}\right)^{-1}
-\left(\sqrt{1-\alpha^2hW^2}\right)^{-1}\right]^2\nonumber\\
& &+\frac{1}{2}e^{-\delta\varphi}
\sqrt{1-fG^{ab}\partial_aW\partial_bW}\sqrt{1-{e^{-2\gamma\varphi}}fG_{ab}
\dot{\phi}^a\dot{\phi}^b}\nonumber \\
& &\times G_{cd}\left[\left(\sqrt{1-{e^{-2\gamma\varphi}}fG_{ab}
\dot{\phi}^a\dot{\phi}^b}\right)^{-1}e^{\delta\varphi}\dot{\phi}^c
+\left(\sqrt{1-fG^{ab}\partial_aW\partial_bW}\right)^{-1}\epsilon
G^{ce}\partial_e\mathcal{W}\right]\nonumber \\ 
& &\qquad\times\left[\left(\sqrt{1-{e^{-2\gamma\varphi}}fG_{ab}
\dot{\phi}^a\dot{\phi}^b}\right)^{-1}e^{\delta\varphi}\dot{\phi}^d
+\left(\sqrt{1-fG^{ab}\partial_aW\partial_bW}\right)^{-1}\epsilon
G^{de}\partial_e\mathcal{W}\right]\nonumber \\
& &-\frac{1}{2f}e^{(2\gamma+\delta)\varphi}
\sqrt{1-fG^{ab}\partial_aW\partial_bW}\sqrt{1-{e^{-2\gamma\varphi}}fG_{ab}
\dot{\phi}^a\dot{\phi}^b}\nonumber \\
& &\qquad\times \left[\left(\sqrt{1-{e^{-2\gamma\varphi}}fG_{ab}
\dot{\phi}^a\dot{\phi}^b}\right)^{-1}-
\left(\sqrt{1-fG^{ab}\partial_aW\partial_bW}\right)^{-1}\right]^2\nonumber \\ 
& &-\epsilon(\dot{\varphi}\partial_\varphi\mathcal{W}+
\dot{\phi}^a\partial_a\mathcal{W})\,.
\end{eqnarray}
Therefore, the equations (\ref{BPSlast}) with (\ref{lastP}) 
correspond to a stationary point of the action.
Note that due to (\ref{eq1}) and (\ref{eq2}), the BPS-type equations becomes
\begin{equation}
e^{\delta\varphi}\dot{\varphi}=\epsilon\partial_\varphi\mathcal{W}(\varphi,\phi)\,,\quad
e^{\delta\varphi}\dot{\phi}^a=-\epsilon
G^{ab}\partial_b\mathcal{W}(\varphi,\phi)\,,
\label{BPSlast}
\end{equation}
which is independent of $f(\phi)$ and $h(\phi)$, i.e., these are the same as the
equations for the canonical theory obtained in the limit $f(\phi)\rightarrow 0$,
$h(\phi)\rightarrow 0$.

Thus, two slow-roll parameters are expressed, when we consider the
single-field case as
\begin{equation}
\varepsilon_H=-\frac{\dot{H}}{H^2}=2(D-2)\frac{W'(\phi)^2}{W(\phi)^2}\,,
\quad
\eta_H=-\frac{\ddot{\phi}}{H\dot{\phi}}=2(D-2)
\frac{W''(\phi)}{W(\phi)}\,.
\end{equation}
We conclude that this model (B) is not very useful for inflation and 
cosmology of the early universe.

\section{Summary and prospects}
\label{conclusion}

In this parer, we propose models of the $n$-DBI gravity coupled to the DBI-type
scalar theory. We showed that the specified scalar potential leads to the
cosmological evolution which is governed by the coupled first-order differential
equations of the BPS-type. We mainly examined analytical investigation on the
cases with the single-field prepotential $W(\phi)$.
We also showed the slow-roll parameters in the case with the single-field
prepotential and with the other constant functions $f(\phi)=f_0$ and
$h(\phi)=h_0$, in compact forms.

As future work, we should study numerical analyses of the models, and the
inclusion of possible matters and radiations.
Russo \cite{Russo2} also considered the general nonlinear sigma model, so,
the possibility of multi-field inflation models is also worth studying in future. 
Moreover, an interesting future challenge will be to search for a model that fits
precision cosmology within that category.

We hope that $n$-DBI gravity and DBI-type theories will also appear in the result
of some kind of operation such as the $T\bar{T}$ deformation, and it would be good
if it had some connection to UV completion when the theory faces quantization.

\section*{Acknowledgments}
This work was supported by YAMAGUCHI UNIVERSITY FUND(M.T.).

\bibliographystyle{apsrev4-1}


\end{document}